\documentclass[amsmath,showpacs,amssymb,floatfix,prd,
onecolumn,superscriptaddress,nofootinbib]{revtex4}

\usepackage{graphicx,subfigure,bm,color,psfrag,hyperref}
\usepackage{amsfonts}
\usepackage{lipsum}
\usepackage{mathtools}
\usepackage{verbatim}
\usepackage[normalem]{ulem}
\usepackage[dvipsnames]{xcolor}
\hypersetup{colorlinks,linkcolor={blue},citecolor={red},urlcolor={teal}}

\usepackage{amsmath}
\usepackage{multirow}
 \usepackage{soul}
 \usepackage{dcolumn}
\usepackage{amssymb}
 \usepackage{amsbsy}
 \usepackage{rotating}
\usepackage{overpic}
\usepackage{ifxetex}
\usepackage{lipsum}
\usepackage{float}
\usepackage{bbold}
\usepackage{enumerate}
\usepackage{dsfont}
\usepackage{mathrsfs}
\usepackage{mathtools}
\usepackage{tensor}
\usepackage{caption}
\usepackage{subcaption}
\usepackage{ragged2e}

\definecolor{amaranth}{rgb}{0.9, 0.17, 0.31}
\definecolor{palatinateblue}{rgb}{0.15, 0.23, 0.89}
\definecolor{brightpink}{rgb}{1.0, 0.0, 0.5}

\hypersetup{ linktoc=all,
    colorlinks, linkcolor={palatinateblue},
    citecolor={brightpink}, urlcolor={amaranth}
}

\graphicspath{ {images/} }

 \newcommand{\be}{\begin{equation}}
\newcommand{\ee}{\end{equation}}
\newcommand{\ba}{\begin{eqnarray}}
\newcommand{\ea}{\end{eqnarray}}

\def\doi{http://doi.org}

\begin{document}

\title{Non-singular bounce solutions in   Myrzakulov   $f(R,T)$ gravity}

\author{Andreas Lymperis}
\email{alymperis@upatras.gr}
\affiliation{Department of Physics, University of Patras, 26500 Patras, 
Greece}

\author{Gulgassyl Nugmanova} 
\affiliation{Eurasian National University, Astana, 010008, Kazakhstan }

 \author{Almira Sergazina}
\affiliation{Eurasian National University, Astana, 010008, Kazakhstan }


\begin{abstract} 
We investigate the realization of a nonsingular bounce within the framework of 
Myrzakulov \( F(R,T) \) gravity. This modified gravitational theory uses a 
non-special connection that  combines both curvature and torsion, giving rise to 
an effective sector that can easily satisfy the violation of the null energy 
condition. We suitably choose the functions that parametrize the connection in 
order to be able to produce     simple and   matter bounce scale factors  at 
the background level. Finally, we examine the evolution of scalar 
perturbations through the bounce using the Mukhanov-Sasaki formalism, and we 
calculate the power spectrum.

\end{abstract}
\maketitle

\renewcommand{\tocname}{Index}

\section{Introduction}

The standard cosmological model, based on General Relativity (GR) and the 
$\Lambda$CDM scenario, successfully describes the evolution and the  
large-scale structure of the universe. Observations from the Cosmic Microwave 
Background (CMB) \cite{Planck:2018vyg}, Baryon Acoustic Oscillations (BAO) 
\cite{Alam:2016hwk}, and Type Ia Supernovae (SNIa) 
\cite{Riess:1998cb,Perlmutter:1998np} provide strong support for this 
paradigm. However, the model faces significant challenges, particularly in the 
very early times, where it predicts the existence of an initial singularity, 
  \cite{Hawking:1973uf}, where the energy density and spacetime curvature 
diverge, signaling the breakdown of 
GR and the necessity for alternative approaches that can provide a non-singular 
description of the Universe origin.

One promising alternative to circumvent the singularity problem is 
the framework of bouncing cosmology. The bounce realization postulates a prior 
contraction phase followed by an expansion phase, avoiding the infinite 
densities and curvatures of the Big Bang 
\cite{Novello:2008ra,Brandenberger:2016vhg}. A successful bouncing scenario 
requires the violation of the null energy condition (NEC), which can be 
achieved through modified theories of gravity or exotic matter components. 
Having in mind that modified gravity theories have additional advantages, such 
as improved  renormalizability, efficiency to describe the late-time Universe, 
and alleviation of the cosmological tensions 
\cite{CANTATA:2021ktz,Capozziello:2011et}, proceeding towards this direction is 
more efficient. Hence, in  the literature one can find many bounce realizations 
in various modified gravity theories  \cite{Khoury:2001wf, Khoury:2001bz, 
Nojiri:2006ri,Cai:2010zma,Czuchry:2010vx,Qiu:2011cy,Avelino:2012ue,
Yoshida:2017swb, Biswas:2005qr,Singh:2006im,
Biswas:2006bs,Creminelli:2007aq,Nojiri:2014zqa,Battefeld:2014uga,
Pavlovic:2017umo,Cai:2011tc,Sahoo:2019qbu,Cai:2012ag,Bamba:2014mya,Peter:2002cn,
Shtanov:2002mb,Martin:2003sf, 
Saridakis:2007cf,Arefeva:2007wvo,Cai:2009in,Barragan:2009sq,Bojowald:2001xe, 
Cai:2014zga,Singh:2015bca,Oikonomou:2015qha,
Banerjee:2016hom,Li:2016xjb,
Cai:2014jla, 
Odintsov:2015uca,Nojiri:2017ncd,Lymperis:2017ulc,Qiu:2018nle,Ilyas:2020qja,
Renevey:2020zdj,
Saridakis:2018fth,Ijjas:2018qbo,Nojiri:2019yzg,Bajardi:2020fxh,
Casalino:2020kdr,Barboza:2020jux, Minas:2019urp,Elizalde:2020zcb,
Zhu:2021whu,Bombacigno:2021bpk,Ilyas:2021xsu,Odintsov:2021urx,Ageeva:2022asq, 
Odintsov:2022qnn,Agrawal:2022vdg,Lymperis:2022oyo,Banerjee:2022gpy,
Battista:2022hqn,Asimakis:2022jel,
Alonso-Serrano:2022nmb, Hu:2023ndc, 
Suzuki:2023osp,Saha:2023ziw,Ghosh:2024cab,Jawad:2024leu}.

The construction of modified gravity theories can be obtained by breaking or 
extending the foundations and assumptions of General Relativity. For instance 
one can start from the   Einstein-Hilbert action  and add extra terms, 
resulting 
to   $f(R)$   gravity \cite{DeFelice:2010aj,Nojiri:2010wj,Starobinsky:2007hu},   
Gauss-Bonnet and $f(G)$ gravity    
\cite{Nojiri:2005jg,DeFelice:2008wz,Zhao:2012vta,Shamir:2020ckh},   Lovelock  
gravity \cite{Lovelock:1971yv,Deruelle:1989fj} etc. On the other hand, one can 
start form the   torsional     gravitational framework and obtain
  $f(T)$ gravity 
\cite{Cai:2015emx,Bengochea:2008gz,Linder:2010py,Chen:2010va,Tamanini:2012hg,
Bengochea:2010sg,
Liu:2012fk,Daouda:2012nj},  
$f(T,T_G)$ gravity 
\cite{Kofinas:2014owa,Kofinas:2014daa},   $f(T,B)$ gravity
\cite{Bahamonde:2015zma,Farrugia:2018gyz}, etc.
Additionally, one could use nonmetricity and the symmetric teleparallel 
connection and result to  $f(Q)$ gravity  
\cite{BeltranJimenez:2017tkd,Heisenberg:2023lru}, to $f(Q,C)$ gravity 
\cite{De:2023xua}, etc. Finally, one can extend the above classes of 
modifications by additionally assuming couplings with a scalar field, such as 
in 
   Horndeski gravity and generalized Galileon theory 
\cite{Horndeski:1974wa,DeFelice:2010nf,Deffayet:2011gz}, in teleparallel 
Horndeski \cite{Bahamonde:2019shr}, etc.

In these lines,  Myrzakulov $F(R,T)$ gravity uses both curvature and torsion to 
describe gravity \cite{Myrzakulov:2012qp}. In particular, using  a non-special 
connection that simultaneously possesses both curvature and torsion one can 
obtain a richer geometric structure that can lead to richer cosmological 
phenomenology and behavior, and that is why it has been applied in      various 
contexts, including late-time cosmic acceleration, 
inflationary scenarios, and dark energy models   
\cite{Myrzakulov:2012qp, Saridakis:2019qwt,
Anagnostopoulos:2020lec,
Myrzakul:2021kas,Iosifidis:2021kqo,Myrzakulov:2021vel,Harko:2021tav,
Papagiannopoulos:2022ohv,Kazempour:2023kde,Zahran:2024nsz,Maurya:2024btu,
Maurya:2024nxx,Maurya:2024ign,Maurya:2024mnb,Momeni:2024bhm}.

In this work we are interested in using  Myrzakulov $F(R,T)$ gravity in order 
to obtain bouncing cosmology. Given the additional terms in the gravitational 
field equations due to the interplay between curvature and torsion, it is 
plausible that bouncing solutions can naturally arise. In particular, we will  
adopt the mini-superspace approach to construct explicit bouncing solutions and 
analyze their stability.  
The paper is structured as follows: In Section \ref{section2}, we review the 
fundamentals of Myrzakulov  $F(R,T)$ gravity and its field equations. In 
Section 
\ref{BouncingSolutions} we construct explicit example of bounce realization, 
examining also the violation of the null energy condition. Then, in Section  
\ref{perturbations} we study the scalar perturbations and we extract the power 
spectrum. Finally, conclusions and future prospects are presented in Section 
\ref{Conclusions}.

\section{Myrzakulov $F(R,T)$ Gravity and Cosmology}
\label{section2}

In this section we present a brief review of Myrzakulov $F(R,T)$ gravity,  
extracting the relevant cosmological equations.

\subsection{Myrzakulov Gravity}

The central idea of this modified gravity is 
the modification of the underlying connection. In particular, it is known that 
imposing a general connection $\omega^{a}_{\,\,\,bc}$ one defines the 
 curvature and the torsion tensor respectively 
 as \cite{Saridakis:2019qwt}
\begin{eqnarray}
&&\!\!\!\!\!\!\!\!\! R^{a}_{\,\,\, b\mu\nu}\!=\omega^{a}_{\,\,\,b\nu,\mu}-
\omega^{a}_{\,\,\,b\mu,\nu}
+\omega^{a}_{\,\,\,c\mu}\omega^{c}_{\,\,\,b\nu}-\omega^{a}_{\,\,\,c\nu}
\omega^{c}_{\,\,\,b\mu}\,,
\label{curvaturebastard}
\end{eqnarray}
\begin{equation}
T^{a}_{\,\,\,\mu\nu}=
e^{a}_{\,\,\,\nu,\mu}-e^{a}_{\,\,\,\mu,\nu}+\omega^{a}_{\,\,\,b\mu}e^{b}_{\,\,\,
\nu}
-\omega^{a}_{\,\,\,b\nu}e^{b}_{\,\,\,\mu}\,,
\label{torsionbastard}
\end{equation}
where $e^{\,\,\, \mu}_a\partial_\mu$ is the tetrad field related to the metric 
as  $
g_{\mu\nu} =\eta_{ab}\, e^a_{\,\,\,\mu}  \, e^b_{\,\,\,\nu},
$ with $\ \eta_{ab}=\text{diag}(1,-1,-1,-1)$ (Greek and Latin indices 
denote coordinate and tangent space respectively), and where we have used
comma for differentiation.

 The  Levi-Civita     connection $\Gamma_{abc}$ is the only one that leads to  
vanishing torsion. In the following we use the label ``LC'' to denote the 
 curvature (Riemann) tensor corresponding to $\Gamma_{abc}$:
 $R^{(LC)a}_{\,\,\,\ \ \ \ \ \, 
b\mu\nu}=\Gamma^{a}_{\,\,\,b\nu,\mu}-
\Gamma^{a}_{\,\,\,b\mu,\nu}
+\Gamma^{a}_{\,\,\,c\mu}\Gamma^{c}_{\,\,\,b\nu}-\Gamma^{a}_{\,\,\,c\nu}
\Gamma^{c}_{\,\,\,b\mu}$.
Similarly,  the Weitzenb{\"{o}}ck connection  
$W_{\,\,\,\mu\nu}^{\lambda}=e_{a}^{\,\,\,\lambda}e^{a}_{\,\,\,\mu
, \nu } $ yields   zero curvature, leading   to the torsion tensor
 $T^{(W)\lambda}_{\,\,\,\ \ \ \ \ \mu\nu}=W^{\lambda}_{\,\,\,\nu\mu}-
W^{\lambda}_{\,\,\,\mu\nu}$.  In the following we use the label
  ``W'' to denote quantities arising from the use of
$W_{\,\,\,\mu\nu}^{\lambda}$.
In summary,  the Ricci scalar corresponding to the 
Levi-Civita connection is 
  \begin{eqnarray}
 &&
 \!\!\!\!\!\!\!\!\!\!\!\!\!\!\!\!\!\!\!\!\!\!\!\!\!\!\!\!\!\!
 R^{(LC)}=\eta^{ab} e^{\,\,\, \mu}_a e^{\,\,\, \nu}_b  \left[
 \Gamma^{\lambda}_{\,\,\,\mu\nu,\lambda}
-
 \Gamma^{\lambda}_{\,\,\,\mu\lambda,\nu} 
 + \Gamma^{\rho}_{\,\,\,\mu\nu}\Gamma^{\lambda}_{\,\,\,\lambda\rho}
-\Gamma^{\rho}_{\,\,\,\mu\lambda}\Gamma^{\lambda}_{\,\,\,\nu\rho}
  \right],
\end{eqnarray}
and the torsion scalar corresponding to the Weitzenb{\"{o}}ck connection 
is expressed as
\begin{eqnarray} 
&&
\!\!\!\!\!\!\!\!\!\!
T^{(W)}=\frac{1}{4}
\left(W^{\mu\lambda\nu}-
W^{\mu\nu\lambda} \right)
\left(W_{\mu\lambda\nu}  -W_{\mu\nu\lambda}\right) 
+\frac{1}{2} \left(W^{
\mu\lambda\nu }
-W^{
\mu\nu\lambda } \right)
\left(W_{\lambda\mu\nu}
-W_{\lambda\nu\mu}\right)
\nonumber\\
&&\ \ \ \ \ 
- \left(  
W_{\nu}^{\,\,\,\mu\nu}
-W_{\nu}^{\,\,\,\nu\mu}\right)  
\left( W^{\lambda}_{\,\,\,\mu\lambda}-W^{\lambda}_{\,\,\,\lambda\mu}\right).
\label{TdefW}
 \end{eqnarray} 

In  Myrzakulov   gravity a non-special  connection which has  non-zero 
curvature and torsion is applied \cite{Saridakis:2019qwt}.
Thus, the torsion and curvature scalars calculated with the use of this 
non-special connection can be written as
\begin{eqnarray}
&&
 T=\frac{1}{4}T^{\mu\nu\lambda}T_{\mu\nu\lambda}+\frac{1}{2}T^{\mu\nu\lambda}
T_{\lambda\nu\mu}-T_{\nu}^{\,\,\,\nu\mu}T^{\lambda}_{\,\,\,\lambda\mu},
\label{Tdef2}
\\
&&
 R=R^{(LC)}+T-2T_{\nu\,\,\,\,\,\,\,\,;\mu}^{\,\,\,\nu\mu}\,,
 \label{Radef222}
 \end{eqnarray}
where $;$ denotes  the covariant differentiation in terms of the 
Levi-Civita connection.
Hence, one can now write the action of the theory as  
\begin{equation}
S = \int d^{4}x e \left[ \frac{F(R,T)}{2\kappa^{2}}   +L_m \right],
\label{action1}
\end{equation}
where $e = \text{det}(e_{\mu}^a) = \sqrt{-g}$ is the tetrad determinant,  and
$\kappa^2=8\pi G$ is  the 
gravitational constant. Additionally, $L_m$ is the matter Lagrangian.

 As we can see $R$ depends on the tetrad and its first and second
derivatives, and on the connection and its first derivative,  and  
 $T$ depends on the tetrad field, its first 
derivative and the connection. Hence, one can parametrize them as
 \begin{eqnarray}
&&T=T^{(W)}+v, 
\label{T1}
\\
&&
R=R^{(LC)} + u, 
\label{R1}
\end{eqnarray}
with $v$   a scalar function depending   on the tetrad, its first 
derivative and the connection, and $u$   a scalar function  depending on 
the   tetrad, its first and second derivatives, and the connection 
and its first derivative.

 \subsection{Cosmology}
 
  We proceed to the application in a   flat 
Friedmann-Robertson-Walker (FRW) geometry of the form
\begin{eqnarray}
 ds^2= dt^2-a^2(t)\,  \delta_{ij} dx^i dx^j,
 \end{eqnarray}
 with $a(t)$ is the scale factor,
which arises from   the tetrad   
$e^a_{\,\,\,\mu}={\text{ diag}}[1,a(t),a(t),a(t)]$. 
As usual, we have 
 $R^{(LC)}=6   \left( \frac{\ddot{a}}{a}+  \frac{\dot{a}^{2}}{a^2}\right)$ and 
$T^{(W)}=-6 \left(  \frac{\dot{a}^{2}}{a^2} \right)$.
Following 
\cite{Saridakis:2019qwt}  we apply the mini-superspace  approach. 
Hence, we impose that  $u=u(a,\dot{a},\ddot{a})$ and $v=v(a,\dot{a})$, while 
for the matter Lagrangian we have  $L_m=-\rho_m(a)$, where $\rho_m$ is the 
matter perfect fluid energy density
\cite{Paliathanasis:2015aos}.

Let us now choose the function  $F(R,T)$. In order to keep the basic versions 
of the theory, we choose  $F(R,T)=R+\lambda T$ with $\lambda$ a dimensionless 
parameter. Note that even in this linear case the theory is still rich due to 
the role of the non-special connection, parameterized through $u$ and $v$.
 Inserting  everything
into  (\ref{action1})   we obtain $S=\int 
Ldt$, with 
\begin{eqnarray}
&&
\!\!\!\!\!\!\!\!\!\!\!
L= 
\frac{3}{\kappa^2}\left[\lambda+1\right]a\dot{a}^{2}-
\frac{ a^{3}}{2\kappa^2}\left[ u(a,\dot{a},\ddot{a})+\lambda
v(a,\dot{a}) \right] 
+ a^3  \rho_m(a).
\end{eqnarray}
Thus, performing variation we find \cite{Saridakis:2019qwt} 
\begin{eqnarray}
3H^{2}&=& \kappa^2\left( 
\rho_m+\rho_{MG} \right)
\label{FR1a}
\\
2\dot{H}+3H^2
&=&  -\kappa^2 \left(p_m+ p_{MG}\right),
\label{FR2a}
\end{eqnarray}
where   $H=\frac{\dot{a}}{a}$  is the Hubble parameter and  $p_m$   the matter 
pressure,
which are the two Friedmann equations in the scenario at hand. Note that we 
have defined an effective   energy density and pressure  of gravitational 
origin given as
\begin{eqnarray}
&&
\!\!\!\!\!\!\!\!\!\!\!\!\!\!\!\!
\rho_{MG}=\frac{1}{\kappa^2} \left[
\frac{Ha}{2} \left(u_{\dot{a}}+v_{\dot{a}} \lambda\right) -\frac{1}{2} 
(u+\lambda v) 
+
\frac{a u_{\ddot{a}}}{2}  \left(\dot{H}-2 H^2\right)    
-3\lambda H^2\right]
\label{rhoDEa1}\\
&&
\!\!\!\!\!\!\!\!\!\!\!\!\!\!\!\!
p_{MG}=
-\frac{1}{\kappa^2}
\left[\frac{Ha}{2} \left(u_{\dot{a}}+v_{\dot{a}} 
\lambda\right)
 -\frac{1}{2} (u+ \lambda v)
 -\frac{a}{6} 
\left(u_a+\lambda v_a-{\dot{u}_{\dot{a}}}-\lambda 
{\dot{v}_{\dot{a}}}\right)\right.\nonumber\\
&& \left. \ \  \ \  \ \  \ \,  \ 
-\frac{a}{2}\left(\dot{H}+3H^2\right)u_{\ddot{a}}-H a 
\dot{u}_{\ddot{a}}
-\frac{a}{6} 
\ddot{u}_{\ddot{a}}
-\lambda(2\dot{H}+3H^2)\right],
\label{pDEa1}
\end{eqnarray}
where   the indices 
$a,\dot{a},\ddot{a}$ denote partial derivatives with respect to this 
argument. Finally, from the form of these equations we deduce that the matter 
energy density and the  modified gravity    energy density are 
conserved, namely  
$ \dot{\rho}_m+3H(\rho_m+p_m)=0$
and
\begin{eqnarray}
 \dot{\rho}_{MG}+3H(\rho_{MG}+p_{MG})=0.
\end{eqnarray}

\section{Bouncing Solutions in Myrzakulov $F(R,T)$ Gravity}
\label{BouncingSolutions}

In this section, we investigate bouncing cosmological solutions within the 
framework of Myrzakulov $F(R,T)$ gravity. Firstly, we present the general 
requirements for a bounce realization and then we construct specific 
realizations.

  \subsection{Bounce Cosmology: General Considerations}

A cosmological non-singular bounce is characterized by a transition from a 
contracting phase 
($\dot{a} < 0$) to an expanding phase ($\dot{a} > 0$), implying that the scale 
factor $a(t)$ reaches a minimum value at  a the bounce point $t_B$, where:
\begin{equation}
    \dot{a}(t_B) = 0, \quad \ddot{a}(t_B) > 0.
\end{equation}
This implies that the Hubble parameter exhibits a transition from negative to 
positive, being zero at the bounce point. Hence, for a non-singular bounce, 
the energy density and pressure of the universe 
should remain finite throughout the evolution.  
 
Achieving a viable bounce requires the violation of the Null Energy Condition 
(NEC), which states that for any null vector $k^\mu$, the energy-momentum 
tensor must satisfy:
\begin{equation}
    T_{\mu\nu} k^\mu k^\nu \geq 0.
\end{equation}
In an FRW geometry the NEC simplifies to:
\begin{equation}
    \rho + p \geq 0.
\end{equation}
Thus, for a bounce to occur this condition must be violated near the bounce 
point, and this violation requires either exotic forms of matter or 
gravitational modifications \cite{Cai:2012va}. 
Concerning modified gravity one     may have a
matter bounce, in which    the universe contracts in a matter-dominated phase 
before transitioning to an expanding phase, leading to a nearly scale-invariant 
spectrum of perturbations   
\cite{Brandenberger:2012zb}, an  Ekpyrotic bounce, which  
exhibits a slow contraction driven by a steep potential  which suppresses 
anisotropies and inhomogeneities,   \cite{Khoury:2001wf}, the quantum 
bounce, in which  loop-quantum-cosmology 
 non-perturbative corrections   prevent singularities and 
result in a smooth bounce \cite{Ashtekar:2011ni}, an  $f(T)$ gravity bounce 
\cite{Cai:2011tc}, etc. In summary, a
viable bounce scenario must satisfy conditions that ensure a smooth transition 
from contraction to expansion while being compatible with observational 
constraints.

  \subsection{Bouncing solutions}

Let us now  extract specific bouncing solutions within the Myrzakulov $F(R,T)$ 
gravity. In particular, we will consider a simple bouncing scale factor and 
reconstruct the deformation functions \( u(a, \dot{a}, \ddot{a}) \) 
and \( v(a, \dot{a}) \) to achieve a consistent bounce scenario.

  \subsubsection{Basic bounce}
  
A simple and commonly studied bouncing scale factor that ensures a non-singular 
evolution of the universe is given by the form:
\begin{equation}
    a(t) = a_B \left(1 + \sigma t^2 \right),
    \label{scalefactor}
\end{equation}
where \( a_B \) is the scale factor at the bounce point  and \( \sigma \) is a 
positive constant controlling the bounce duration. This scale factor satisfies 
the bouncing conditions, namely 
  \( \dot{a}(t) = 2\sigma t a_B > 0 \) for \( t > 0 \) (expansion), and 
\( \dot{a}(t) < 0 \) for \( t < 0 \) (contraction). Additionally, it satisfies 
  \( \ddot{a}(t) = 2\sigma a_B > 0 \) at the bounce point \( t = 0 \), 
ensuring a smooth transition from contraction to expansion.
 The Hubble parameter for this scale factor is given by:
\begin{equation}
    H(t) = \frac{\dot{a}}{a} = \frac{2\sigma t}{1 + \sigma t^2},
    \label{Hubble}
\end{equation}
which satisfies \( H(0) = 0 \) and exhibits a transition from negative to 
positive values across the bounce.

Our goal is to reconstruct the deformation functions \( u \) and \( v \) such 
that they satisfy the Friedmann equations under the bounce scale factor.
We proceed by choosing the ansatz of $u(a, \dot{a}, \ddot{a})$ and $ v(a, 
\dot{a})$. We start by the form 
\begin{equation}
    u(a, \dot{a}, \ddot{a}) = \beta \frac{\ddot{a}}{a} + \gamma 
\frac{\dot{a}}{a},
\end{equation}
\begin{equation}
    v(a, \dot{a}) =  \delta \frac{\ddot{a}}{a}+\epsilon \frac{\dot{a}}{a},
\end{equation}
where \( \beta, \gamma, \) and \(\delta, \epsilon \) are constants.
Inserting these forms, alongside (\ref{scalefactor}), into  the Friedmann 
equations
(\ref{FR1a}),(\ref{FR2a}), and considering for simplicity $\lambda=1$, we find 
that they are fulfilled if we consider 
\begin{equation}
 \beta=\frac{24 a_B}{5}; \ \ \ 
 \delta=-\frac{24}{5} (a_B+5 ),
\end{equation}
 with $\gamma$ and $\epsilon$ arbitrary, making the identification 
$\rho_ B\equiv\rho_m(a_B)=\frac{144}{5\kappa^2}\sigma a_B^3$.

  \subsubsection{Matter bounce}
  
 Let us now proceed to the realization of matter bounce. This solution has 
attracted the interest of the community from the time the works
\cite{Wands:1998yp,Finelli:2001sr} appeared, since it can give rise to an 
almost 
scale-invariant power spectrum of 
primordial perturbations 
\cite{Brandenberger:2007by,Cai:2007zv}.

In this case the  
 scale factor is
\begin{equation}
 a(t)=a_{B}\left( 1+\frac{3}{2}\sigma t^{2}\right) ^{1/3},
\label{atbounce}
\end{equation}
since such an ansatz exhibits the bouncing
behavior  corresponding to matter-dominated
contraction and expansion.
In this case  
\begin{equation}
H(t)=\frac{\sigma t}{(1+3\sigma t^{2}/2)}.
\end{equation} 
 We impose the ansatz
\begin{equation}
    u(a, \dot{a}, \ddot{a}) =   \gamma \frac{\dot{a}}{a} ,
\end{equation}
\begin{equation}
    v(a, \dot{a}) =  6\frac{\ddot{a}}{a}+\epsilon \frac{\dot{a}}{a} 
+\frac{\zeta}{a(t)^3},
\end{equation}
where \(  \gamma, \zeta\) and \( \epsilon \) are constants.
As we can see, the matter bounce scale (\ref{atbounce}) satisfies the Friedmann 
equations if we make the identification
$\rho_ B\equiv\rho_m(a_B)= \frac{ \zeta+4\sigma a_B^3}{2 \kappa^2}$.

To ensure the physical viability of the bouncing solution, we analyze the 
energy conditions. For the effective energy density and pressure 
(\ref{rhoDEa1}), (\ref{pDEa1}), we obtain
\begin{eqnarray}
   \kappa^2( \rho_{MG} + p_{MG}) =
 (a u_{\ddot{a}}-2 \lambda)  \dot{H}    
+\frac{a}{6} 
\left(u_a+\lambda v_a-{\dot{u}_{\dot{a}}}-\lambda 
{\dot{v}_{\dot{a}}}+
\ddot{u}_{\ddot{a}}\right) 
+\frac{a}{2}   H^2 u_{\ddot{a}}+H a 
\dot{u}_{\ddot{a}} .
\end{eqnarray}
 As we observe, the  null energy condition (NEC) violation is satisfied for the 
above examples. Furthermore, we deduce that choosing suitable ansätze for $u(a, 
\dot{a}, \ddot{a})$ and $ v(a, 
\dot{a})$ that satisfy     $\rho_{\text{eff}} + p_{\text{eff}} < 0$ then the 
bounce realization is ensured.
 
 \section{Evolution of perturbations through the bounce}
 \label{perturbations}

In this section, we examine the behavior of scalar perturbations through the 
bounce within the framework of Myrzakulov $F(R,T)$ gravity. To investigate the 
dynamics of cosmological perturbations, we use the Mukhanov-Sasaki equation, 
which describes the evolution of scalar perturbations in a perturbed FRW 
universe.

In the presence of a bouncing scale factor $a(t)$, scalar perturbations are 
governed by the Mukhanov-Sasaki equation:
\begin{equation}
v_k'' + \left(c_s^2 k^2 - \frac{z''}{z}\right) v_k = 0,
\label{eq:MS}
\end{equation}
where $v_k$ is the Mukhanov-Sasaki variable, $c_s$ is the sound speed, $k$ is 
the comoving wave number, and $z$ is given by
\begin{equation}
z = \frac{a \sqrt{\rho_{t} + p_t}}{c_s H}.
\end{equation}
Here, $\rho_t$ and $p_t$ denote the effective energy density and pressure, 
which include contributions from both matter and the modified gravity sector.

During the contracting phase, the term $z''/z$ in Eq.~\eqref{eq:MS} dominates 
over $c_s^2 k^2$ for long-wavelength modes ($k \to 0$). As the bounce 
approaches, $z''/z$ exhibits a transition, and the evolution of perturbations 
depends critically on the background quantities  as:
\begin{equation}
\frac{z''}{z} \propto 2 H^2 + \dot{H} + \frac{\ddot{H}}{H} - 
\frac{\dot{H}^2}{H^2}.
\end{equation}
Hence, for the simple bounce (\ref{scalefactor}) we find:
\begin{equation}
\frac{z''}{z} \propto \frac{6\sigma}{(1 + \sigma t^2)^2}.
\end{equation}
 The spectrum of scalar perturbations can be evaluated at late times after the 
bounce. The solution to Eq.~\eqref{eq:MS} in the super-horizon limit ($k \to 0$) 
is given by $
v_k \propto z$.
Matching the solutions across the bounce allows us to evaluate the power 
spectrum:
\begin{equation}
\mathcal{P}_\zeta = \frac{k^3}{2\pi^2} \left| \frac{v_k}{z} \right|^2 \propto 
k^{n_s - 1},
\end{equation}
where $n_s$ is the spectral index.   

To facilitate a detailed analysis in the case of matter bounce (\ref{atbounce}) 
we  use that during the  matter-dominated contracting phase, the scale factor 
evolves as
$a \propto t^{2/3} \propto \tau^2$, where $\tau$ is the comoving time 
$\tau\equiv\int dt/a$, 
with \( z \propto a \). In this phase, solving the background equations of 
motion yields approximate expressions for the Hubble parameter  as
\begin{equation} 
H \simeq \frac{2}{3t},  
\end{equation}
which are valid long before the onset of the bouncing phase. Furthermore, 
during this period, the sound speed of curvature perturbations approaches \( 
c_s^2 \simeq 1 \), consistent with the matter-like contracting regime.

As a result, the perturbation equation in the contracting phase takes the form:
\begin{equation}
v_{k}^{\prime \prime }+\left( k^{2}-\frac{2}{\tau ^{2}}\right) v_{k} \simeq 0,  
\label{v_eom_m}
\end{equation}
where the first term \(k^2\) initially dominates. In this regime, the 
gravitational term can be neglected, reducing the equation to that of a free 
scalar field propagating in a flat spacetime. Consequently, the fluctuations are 
well-approximated by the Bunch-Davies vacuum state, which is expressed as:
$
v_{k}\simeq \frac{e^{-ik\tau }}{\sqrt{2k}}$ (setting the sound speed to 1).
Using this   initial condition, we   solve the perturbation equation 
and find:
\begin{equation}
v_{k}=\frac{e^{-ik\tau }}{\sqrt{2k}}\left(1-\frac{i}{k\tau}\right)~.
\end{equation}
This solution demonstrates that quantum fluctuations can evolve into classical 
perturbations as they exit the Hubble radius. This transition occurs due to the 
influence of the gravitational term in Eq.~(\ref{v_eom_m}). Finally, the 
amplitude of these metric perturbations continues to grow until the universe 
enters the bouncing phase.
Hence, we finally find  the expression for the primordial power spectrum 
as
\begin{eqnarray}
P_{\zeta} \equiv \frac{k^3}{2\pi^2}\left|\frac{v_k}{z}\right|^2 =
\frac{\sigma    G}{ 36 \pi  }~.
\end{eqnarray}

In summary, using the  Mukhanov-Sasaki formalism we showed that the bounce is 
stable and free from singularities for appropriately chosen parameters. 
Additionally, the presence of a well-behaved $z''/z$ term ensures that the 
scalar perturbations remain finite throughout the bounce.

\section{Conclusions}
\label{Conclusions}

 Myrzakulov \( F(R,T) \) gravity incorporates both curvature and torsion as 
fundamental components in the description of gravity. By employing a generalized 
connection that simultaneously exhibits both curvature and torsion, this 
framework achieves a more intricate geometric structure. Such an   
geometry opens the way to a wider range of cosmological phenomena,  and thus it 
has been widely applied in diverse contexts.
On the other hand,  bouncing cosmology may act as an alternative to standard 
cosmological paradigm, free from the  singularity problem. The bounce 
realization postulates a prior contraction phase followed by an expansion phase, 
and thus a successful bouncing scenario requires the violation of the null 
energy condition. Since this can be easily obtained in modified gravity 
scenarios, in the literature one may find many bounce realizations in the 
framework of alternative gravitational theories.

In this work we investigated in detail the bounce realization within  
Myrzakulov \( F(R,T) \) gravity. Our results focused on the detailed background 
realization of bouncing solutions and the evolution of perturbations through the 
bounce. In particular, we focused on the realization of matter bounce, since it 
can naturally provide the matter expansion phase after the bounce. As we showed, 
by suitably choosing the connection parametrization functions $u$ and $v$, we 
were able to obtain the matter bounce as a background behavior.

We proceeded to the investigation of perturbations, using the 
Mukhanov-Sasaki formalism. In particular,  we analyzed the stability and the
behavior of scalar perturbations, and we confirmed the scale-invariant nature 
of the primordial power spectrum. Additionally, we showed that the quantum 
fluctuations evolve into classical perturbations as they exit the Hubble radius, 
with the perturbations remaining finite throughout the bounce. These findings 
show the viability of Myrzakulov \( F(R, T) \) gravity as a candidate for 
describing the early universe and generating predictions that align with 
observational constraints.

There are several avenues that are interesting for further exploration. One 
could extend the analysis to include tensor and vector perturbations, in order 
to  obtain information on  the full spectrum of gravitational 
waves and their possible observational signatures. Moreover, the coupling 
between curvature, torsion, and matter fields within this framework could be 
further analyzed. Finally, one could study the cases of more complicated \( F(R, 
T) \) forms. These investigations lie beyond the scope of the present work and 
are left for future projects.

\end{document}